\documentclass{statsoc}

\usepackage[a4paper]{geometry}
\usepackage{graphics,graphicx}
\usepackage[textwidth=8em,textsize=small]{todonotes}
\usepackage{amsmath}
\usepackage{amsfonts}
\usepackage{color}
\usepackage[portuges,english]{babel}
\usepackage[latin1]{inputenc}
\usepackage{placeins}

\title[Unemployment estimation]{Unemployment estimation: Spatial point referenced methods and models}

\author[Pereira, S. {\it et al.}]{Soraia Pereira}

\address{FCUL, Universidade de Lisboa, Lisboa, Portugal.}

\email{sapereira@fc.ul.pt}

\author{K F Turkman}

\address{FCUL,Universidade de Lisboa, Lisboa, Portugal.}

\author{Luís Correia}

\address{Instituto Nacional de Estatística, Lisboa, Portugal.}

\author[S. Pereira, K F Turkman, L. Correia, H. Rue ]{H{\aa}vard Rue}

\address{King Abdullah University of Science and Technology, Saudi Arabia}

\begin{document}

\maketitle % Insert title
\newcommand{\bbeta}{\boldsymbol{beta}}

\newcommand{\bz}{{\bf z}}
\newcommand{\bx}{{\bf x}}
\newcommand{\bs}{{\bf s}}
\newcommand{\bw}{{\bf w}}
\newcommand{\bW}{{\bf W}}
\newcommand{\by}{{\bf y}}
\newcommand{\btheta}{{\boldsymbol\theta}}

\begin{abstract}
\normalfont {\noindent
Portuguese Labor force survey, from 4th quarter of 2014 onwards, started geo-referencing the sampling units, namely the  dwellings in which the surveys are carried.  This opens  new possibilities  in analysing and estimating   unemployment   and its spatial distribution across any region.   The labor force survey choose, according to an preestablished sampling criteria,  a certain number of dwellings across the nation and survey the number of unemployed in these dwellings. Based on this survey, the National Statistical Institute of Portugal presently uses direct estimation methods to estimate the national unemployment figures. Recently, there has been increased  interest  in estimating these  figures in smaller areas. Direct estimation methods, due to reduced sampling sizes in small areas, tend to produce fairly large sampling variations therefore model based methods, which tend to ``borrow strength'' from area to area by making use of the areal dependence, should  be favored. These model based methods tend use areal counting processes as models and typically introduce spatial dependence through the model parameters by  a latent random effect.

In this paper, we suggest modeling the spatial distribution of residential buildings across Portugal by a Log Gaussian Cox process and the  number of unemployed per residential unit as a  mark attached to these random points. Thus the main focus of the study is to model the spatial intensity  function of  this marked point process.  Number of unemployed in any region can then be estimated using a proper functional of this marked point process. The principal objective of this point referenced method for unemployment estimation is to get reliable estimates at higher  spatial resolutions and  at the same time  incorporate  in the model  the  auxiliary information available at residential units such as average income or education level of individuals surveyed in these units.
  %The comparison of methods based on direct estimation and marked point processes is made.
%particularly for  28 NUTS III regions of Portugal.

\noindent }

\end{abstract}

\section{Introduction}\label{int}

\begin{figure}[ht]
\centering
\includegraphics[width=10cm]{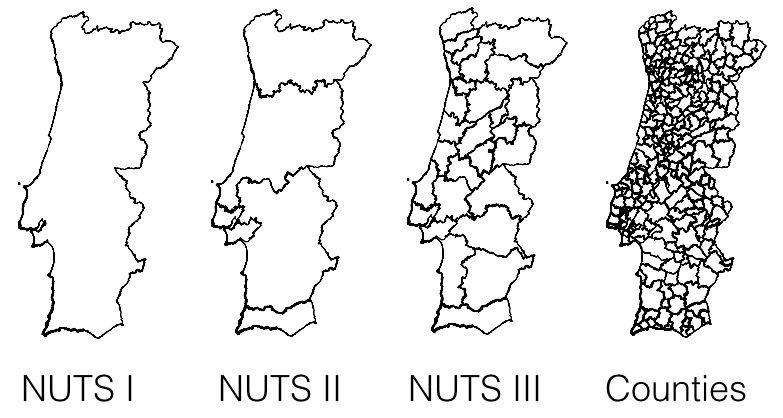}
\caption{NUTS and counties in mainland Portugal}
\label{nuts}
\end{figure}

The knowledge and understanding of unemployment  at a regional level has been increasingly used to make political decisions. In Portugal, the official unemployment figures are published quarterly by the National Statistical Institute of Portugal (INE) at  the national level as well as  for NUTS II regions. NUTS is the classification of territorial units for statistics (see figure \ref{nuts} for a better understanding of NUTS regions and the 278 counties in mainland Portugal). The calculation of the official numbers is based on a direct method from the sample of the Portuguese Labor Force Survey. This method is an extension of the Horvitz-Thompson estimator (Horvitz and Thompson, 1952), with a correction for non-response and a calibration for the known population totals.

Currently, there is a need to obtain reliable estimates at a more disaggregated level, particularly at the NUTS III level. However, due to the relatively small size of these areas, there is insufficient information to obtain estimates with an acceptable level of accuracy using the direct method. Small area estimation methods (Rao, 2003) provide useful tools for such studies. There have been considerable methodological developments to solve small area estimation problems in an unemployment context. See for example, Lopez-Vizcaino \textit{et al} (2015), which proposes a multinomial model for the estimation of the labor force indicators in a classical context, and Pereira \textit{et al} (2016) which proposes alternative spatio-temporal models within a Bayesian context. The majority of small area methods are based on generalized linear models applied to areal data by modeling an appropriate counting process. These methods can "borrow strength" from area to area and make use of auxiliary information at regional level, compensating for the small sample sizes in each area due to the designed sampling survey.

%Portuguese Labor force survey (LFS), from 4th quarter of 2014 onwards, started geo-referencing the sampled units in which the surveys are carried. The adopted  survey uses  dwellings as the sampling unit.  However,  residential units (residential buildings) are geo-referenced and each residential unit   may be composed of one or multiple dwellings.  Thus, in these surveys, different sampling units may have the same spatial location, causing some difficulty in modeling strategies. We will address this problem by moving from dwellings to residential buildings as our sampling units. This strategy will solve the awkward problem of multiplicity of geo-referenced units at the cost of introducing some approximations  at residential unit level with some consequent loss of precision.

From 4th quarter of 2014 onwards all the sampling units (dwellings) of the LFS are geo-referenced according to the coordinates of the respective building, which may be composed by
of one or more dwellings. Thus, different sampling units may have the same spatial location, causing some difficulty in modeling strategies. We will address this problem by moving from dwellings to residential buildings as our sampling units. This strategy will solve the awkward problem of multiplicity of geo-referenced units at the cost of introducing some approximations at residential unit level with some consequent loss of precision.

Geo-referencing  residential units permits using methods and models for point referenced data, increasing  the spatial resolution of inferential methods giving us a much detailed  information on the intensity of unemployment across space.  This approach allows for the representation of the sample survey as a realization of a spatial point process together with  the associated marks, namely the   number of unemployed people in each residential unit. Point referencing also permits us to use more precise auxiliary  information at residential units, such as the average education level or income of individuals living in the same  residential unit.

For modelling the intensity of residential unit  locations and their associated marks, we suggest a marked  Log Gaussian Cox processes (LGCP) as model. The LGCP is a class of flexible models widely used in the context of spatial point processes (Moller and Waagepetersen, 2004, Illian {\it et al}, 2008, Baddeley {\it et al}, 2016). Typically, in this frame work, the log intensity of the point process is assumed to be a  (latent) Gaussian random field. In order to facilitate calculations, often marks are assumed to be independent of point patterns so that marks and spatial patterns can be modeled separately. However, for inference on such models,  Illian  {\it et al} (2012) proposed a flexible framework using INLA (Rue {\it et al}, 2009, Martins {\it et al}, 2013, Rue {\it et al}, 2017), in which  the spatial pattern of points and marks are allowed to be dependent, assuming  independence  conditional on a common latent spatial Gaussian processes, making these models  more flexible.  Inference on such models is not straightforward. Due to  computational problems that emerge in  this framework, Lindgren {\it et al} (2011) proposed a more computationally tractable approach based on stochastic partial differential equation (SPDE) models, which permit the transformation of a Gaussian field  to a Gaussian markov random field and  we follow this method.

The structure of the paper is as follows: In sections \ref{survey} and \ref{data}, we explain the sampling design of the LFS and the consequent data available for the analysis. In section  \ref{spp}, we explain our models as well as inferential methods and give the unemployment estimates for NUTS III regions, as well as for the 278 counties in mainland Portugal. Comparison of reported results with the  direct estimation methods is also given in section  \ref{spp}. Finally in section \ref{discussion}, we give a brief account of possible extensions on models and methods employed.

\section{Labor Force Surveys }\label{survey}

The methodologies proposed in this study are highly dependent on the sampling design of the LFS. Therefore, it is important to understand both how the sampling units are drawn and how the inclusion probabilities are calculated.

The LFS is a continuous survey of the population living in private dwellings within the Portuguese national territory. The survey provides an understanding of the socioeconomic situation of these individuals during the week prior to the interview (reference week). The dwellings are the sampling units and the inhabitants living in these dwellings are the observation units.

The unemployment figures are published quarterly by INE at both the national and NUTS II level. From one quarter to another, the sample changes. These samples have 6 sub-samples, with the oldest one being replaced for a new one in each quarter. This process is also known as a \textit{rotation scheme}. In this way, each individual in the sample is surveyed over 6 consecutive quarters, inducing strong temporal dependence between the quarterly surveys.

%The methodologies proposed in this study are highly dependent on the sampling scheme of the Labor Force Survey (LFS). Therefore, it is  important to understand both how the sampling units  are selected, and how the selection probabilities are calculated.

%The LFS is a continuous survey of all the persons living in private dwellings as usual residence within the Portuguese national territory. The survey provides an understanding of the socioeconomic situation of these individuals during the week prior to the interview. The dwellings are the sampling  units, and the individuals  that reside in these dwellings are the observation units.

%The indicators of the LFS, in the form of unemployment figures, are published quarterly by the INE at both the national level and for NUTS II regions.

Between 2011 and the 2nd quarter of 2013,  samples  were selected from a sampling base called ``master-sample" (MS). From the 3rd quarter of 2013 however, each new sub-sample (in each quarter) was extracted from the national dwellings register (NDR). Contrary to the MS, in NDR all the dwellings are geo-referenced. This transition process from the MS to the NDR was completed in the 3rd quarter of 2014. After the 4th quarter of 2014,  all  dwellings were extracted from the NDR, and therefore, new methodologies based on point-referenced data can  be used. Since there may be more than one dwelling in each residential unit, particularly  in areas of high population density, multiple dwellings in the survey have the same spatial location.

The sampling of the Portuguese LFS is stratified into 30 NUTS III regions. In each region, multi-stage sampling is conducted, where the primary sampling units are areas consisting of one or more cells of the $\text{km}^2$ INSPIRE grid, and the secondary units are dwellings. Every selected dwelling and all its inhabitants are surveyed.
For the selection of the primary units, the dwellings are sorted by their coordinates and a fixed interval $K_h$ is calculated by $K_h=\frac{A_h}{n_h}$, where $A_h$ is the total number of dwellings in strata h and $n_h$ is the  selected number of dwellings  in strata h. A random  variable $u_h \sim U[1,K_h]$ is generated, to determine the position of the dwelling which determines the first area selected (the area in which the dwelling is included). The position of the other selected areas is fixed and determined by the position of the dwelling that determined the position of the previous area plus the interval $K_h$. After the selection of the areas, a similar process is followed to select the dwellings. In each selected area j, a random uniform variable $u_{jh} \sim U[1,K_{jh}]$, is generated, where $K_{jh}=\frac{A_{jh}}{n_{jh}}$,  $A_{jh}$ is the total number of dwellings in the area $j$ and strata $h$ and  $n_{jh}$ is the number of dwellings sampled in area $j$ and strata $h$. $u_{jh}$ defines the first selected dwelling in area $j$ and strata $h$. After the selection of the first dwelling in area $j$, the position of the other dwellings is fixed and determined by the position of the previous selected dwelling plus the interval $K_{jh}$. Following this scheme, the selection probability of the area $j$ in strata $h$ is

$$p_{jh}=\begin{cases}
\frac{A_{jh}}{A_h} \times s_h,&\quad A_{jh} < K_h,\\
1, &\quad otherwise.
\end{cases}
$$

\noindent where $s_h$ is the number of selected areas in the strata h. The selection probability of each dwelling $i$ in area $j$ and strata $h$ is given by

\begin{equation}
p_{ijh}=p_{jh} \times p_{i|jh} = p_{jh} \times \frac{n_{jh}}{A_{jh}},\label{thinning2}
\end{equation}

\noindent Since all the individuals in each selected dwelling are surveyed, their selection probabilities are equal to the respective dwelling, $p_{kijh}=p_{ijh}$ for each individual $k$ in dwelling $i$.

The official estimates of the unemployment figures are calculated using a direct method, based on the Horvitz-Thompson estimator (Horvitz and Thompson, 1952).

\section{Data}\label{data}

In this study, we analyze the quarterly data of the Labor Force Survey (LFS) regarding to the 4th quarter of 2014. The sample size is about 40,000 observations and each individual can be classified into one of the following three categories: employed, unemployed and inactive. Covariate information about the individuals are available, such as gender, age and education level.

Within a point process modelling scheme, the choice of dwellings as sampling units, creates problems. Since residential units are geo-referenced, multiple sampling units appear with the same spatial location. The sampling design and the consequent data are not sufficiently detailed to obtain good information on the multiplicity distribution of dwellings in each residential building, therefore we use residential buildings as design units. Thus we aggregate the number of unemployed observed in dwellings with the same spatial location and we denote by
  $Y(s_j)$,  the number of unemployed in the residential building at spatial position $s_j$.

\subsection{Covariates}\label{covariates}

One of the great benefits of point referenced models for spatial variation of unemployment is that very detailed covariate information can be given at residential building level, like average education and age of the individuals in each residential building. For the locations we have available only one covariate, the population density, which we will use as offset in the model.

\begin{figure}
\centering
\includegraphics[width=1\linewidth]{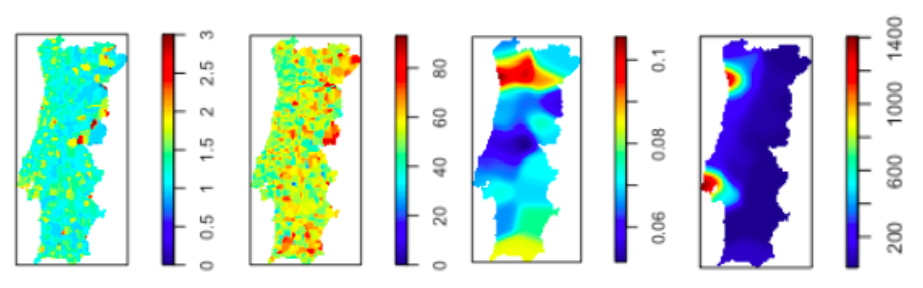}
\caption{Kernel estimates of the median of the education level per residential building, mean age per residential building,  proportion of registered unemployed people in centers of employment and population density.}
\label{covariates2}
\end{figure}

The median of the education level in each residential building and the mean age were considered as covariates to model the marks. Although the education level does not constitute a quantitative variable, it was treated as such due to its ordinal meaning (1-primary level, 2-secondary level, 3-higher level). Higher values of this variable in the Lisbon and Península de Setúbal regions can be  clearly seen  (figure \ref{covariates2}). It is also interesting to see the spatial distribution of the mean age, with more younger people near the country's coastline than in its interior. The proportion of unemployed people registered in the employment centers depends on the number of unemployed and this information is also used as a covariate. Kernel based estimates of the covariates are given in figure \ref{covariates2}.

%The kernel estimates were obtained for three levels of smoothing: k=2, k=10 and k=20. Here, we will use the most detailed level (k=2) due to the specific aim of this study. However, for the proportion of the registered unemployed people, we chose k=20 because of the small number of points in which the estimation was based.

\section{A spatial point patterns approach}\label{spp}

\newcommand{\bZ}{{\bf Z}}

We model the spatial  point process $N_1(\cdot)$  of residential units by  a log  Gaussian Cox process with intensity $\lambda_1(s)$, with
\begin{equation}
\log \lambda_1(s|W(s))= \alpha_1+z'_1(s)\theta_1+W(s),\label{4}
\end{equation}

 We assume that for every $s$, the mark  $Y(s)$  is a  Poisson   random variable  with  probability mass function
\begin{equation}
 P_{Y(s)|W(s)}(y)\sim \text{Poisson}(\lambda_2(s| W(s))),\label{4a}
 \end{equation}
  where
 \begin{equation}
 \log\lambda_2(s|  W(s))=\alpha_2 +z'_2(s)\theta_2+ \alpha_3 W(s).\label{4b}
 \end{equation}
 Here, $W(s)$ is a latent Gaussian Markov field,  $z_1(s)$, $z_2(s)$ are generic auxiliary information which may help in understanding the spatial patterns of points as well as the marks and $\theta=(\alpha_1,\alpha_2,\alpha_3,\theta_1,\theta_2)$ are the model parameters.
  We assume that the same latent Gaussian process $W(s)$ is acting both on the point patterns  and their marks with different scaling and we assume that conditional on $W(s)$,    point mass density and the marks  are independent. It is natural to expect that in areas with high density of residential buildings, one would expect higher rate of unemployment and therefore independence of points and marks may not be an reasonable assumption.

 %Conditional independence relaxes this assumption.  We also consider other  alternative latent dependence structures. For example, as an alternative  model,  we assume that there are two linearly dependent but separate latent Gaussian processes acting on the points and the marks. Similarly in another model, we used two independent gaussian processes for the spatial points and the respective marks. These models are then compared. In this section, we report the model with unique Gaussian structure both for the points and the marks, as it involves simpler notation. Extension of the results to alternative models are straightforward.

With the conditonional independence assumption, the corresponding marked point process $N(s,y)$ has the following structure:
 \begin{equation}
N_1(s,y)|\lambda (s,y) \sim \text{Poisson}(\lambda(s,y)),\label{4d}
\end{equation}
 with
\begin{equation}
\lambda(s,y|W(s))=\lambda_1(s|W(s))P_{Y(s)|W(s)}(y) \label{4c}
\end{equation}

where  $\lambda_1(s|W(s))$, and $P_{Y(s)|W(s)}(y)$ defined as in (\ref{4}),(\ref{4a}) and (\ref{4b}).

%\FloatBarrier

\subsection{Target quantities for inference}\label{targets}

Our objective is to make inference on the number of unemployed in any given region $A$, based on the sample survey. Thus, our target quantity  is a functional of the marked point process, namely  the number of unemployed people in region $A$, given by
 $$N(A) =\sum_{j=1} ^{N_1(A)}  Y(s_j).\label{01}$$

Let $s =(s_1,...,s_n)$ be the location of sampling units chosen in the sampling survey,  $y(s)$ the number of unemployed in each residential unit and $z_1$, $z_2$
the covariates specific to residential units and marks respectively.
We denote by  $\bx=(n,s, y(s), z_1,z_2)$ the observed data obtained from the sampling survey. Our specific target quantities  are  the posterior predictive mean  and variance of the random variable $N(A)$ given by respectively
\begin {eqnarray}
\text{E}(N(A)|\bx)&=& \text{E}_{(W(s), \theta|\bx)}[\text{E}(N(A)|\bx,W(s),\theta)]\nonumber\\
&=& \int_{W(s),\theta} \text{E}(N(A)|\bx,W(s), \theta) p(W(s),\theta |\bx) dW(s)d\theta,  \label{7}
\end{eqnarray}

and

\begin{eqnarray}
\text{Var}(N(A)|\bx)
&=& \text{Var}_{(W(s), \theta|\bx)}\left[\text{E}(N(A)|\bx,W(s), \theta)\right]\nonumber\\
&+&
\text{E}_{(W(s), \theta|\bx)}\left[\text{Var}(N(A)|\bx,W(s), \theta)\right]. \label{7a}
\end{eqnarray}

Calculation of
\begin{equation}
\text{E}(N(A)|\bx,W(s), \theta)=\text{E}(\sum_{j=1}^{N_1(A)}Y(s_j)|\bx,W(s), \theta)
\end{equation}
\noindent and
\begin{equation}
\text{Var}(N(A)|\bx,W(s), \theta)=\text{Var}(\sum_{j=1}^{N_1(A)}Y(s_j)|\bx,W(s), \theta),
\end{equation}
\noindent require certain assumptions.
 \begin{enumerate}

 \item Conditional on $W(s)$,  the point patterns of the Cox process over disjoint regions are independent. Consequently, conditional on $W(s)$, the point patterns over the design pixels $I_{j}$ are  also independent and we also assume that within  each pixel the intensity function of the Cox process is homogeneous so that $\lambda_1(s)=\lambda_1(I_{j})$ for every $s\in I_{j}$.
     \item We assume that conditional on $W(s)$,  the marks $Y(s)$ are independent of the point patterns so that the conditional intensity function of the marked point process is given by
         $$\lambda(s, y| \bx,W(s),\theta)= \lambda_1 (s| \bx,W(s), \theta) P_{Y(s)|bx,W(s), \theta}(y),$$
         \item We assume that conditional on $W(s)$,  marks  observed on disjoint sets are independent.
         \item Finally,  we assume that  the marks $Y(s_i)$  are identical for every
         $s_i\in I_{j}$, that is the number of unemployed in every residential  unit in pixel $I_{j}$ are identical. Hence, in each pixel we replace $\text{E}(Y(s_i))$ by $\text{E}(Y(I_{j}))$.

         \end{enumerate}

To summarize, we  have two major assumptions in this model: The latent Gaussian field $W(s)$ is the only source of dependence in the model. Not only are the point patterns and marks independent conditional on $W(s)$, but the point pattern and marks are independent over disjoint intervals conditional on $W(s)$. Further, within a $\text{km}^2$ design unit pixels, we assume homogeneity of the point patterns as well as marks.
  Let $N(I_{j})$ be the  number of residential units  in each pixel $I_{j}$. Then with assumptions (a)-(e),

\begin{eqnarray}
 \text{E}(N(A)|\bx, W(s),\theta))&=&
 \text{E}(\sum_{j=1}^{N_1(A)}Y(s_j)|\bx,W(s), \theta)\nonumber \\
 &=& \text{E}(\sum_{I_{j}\in A}\sum_{i\in I_{j}} Y(s_i)|\bx,W(s), \theta) \nonumber\\
 &=& \sum_{I_{j}\in A} \text{E}(\sum_{i=1}^{N(I_{j})} Y(s_i)|\bx,W(s), \theta))\\
 &=&\sum_{I_{j}\in A}\text{E}(N_{I_{j}}|W(I_{j}))\text{E}(Y(I_{j})|\bx, W(I_{j}), \theta)   \label{10a}\\
 &=&\sum_{I_{j}\in A} ||I_{j}||\lambda_1(I_{j}|\bx, W(I_{j}),\theta)\lambda_2(I_{j}|\bx,W(I_{j}),\theta)  \label{10b} \\
 &\sim& \int_{s\in A} \lambda_1(s|\bx,W(s),\theta) \lambda_2(s|\bx,W(s),\theta) ds \label{10c}
 \end{eqnarray}

Here, $W(I_{j})$ represents the latent gaussian Markov random field approximating the latent Gaussian random field $W(s)$ obtained by the SPDE method. $w(I_{j})$ values are obtained from INLA as explained  in section \ref{inla}. (\ref{10b}) follows from the conditional independence and  homogeneity of the point patterns as well as the marks within each $\text{km}^2$ pixels, whereas  (\ref{10c}) follows from the approximation of integrals by sums over the design pixels as is explained in \ref{inla} . Thus the $\text{km}^2$ design pixels are the smallest units over which we approximate the point referenced process.

We can calculate, with similar arguments
   \begin{eqnarray}
   \text{Var}(N(A)|\bx,W(s), \theta ) &\sim& \int_{s\in A} \lambda_1(s|\bx,W(s)),\theta) \lambda_2(s|\bx,W(s),\theta) ds \nonumber\\
  &+& \int_{s\in A} \lambda_1(s|\bx,W(s),\theta) \lambda_2^2(s|\bx,W(s),\theta) ds
   \end{eqnarray}
  The  mean and the variance of of the predictive distribution given in (\ref{7}) and (\ref{7a}) can be calculated numerically.
INLA package permits the calculation of the intensity function $\lambda_1(s|\bx, W(s),\theta)$ as well as the mean mark $\lambda_2 \left(s| \bx,W(s),\theta \right)$. INLA also simulates  from the  marginal posterior densities of the latent process as well as the model parameters, thus target quantities  (\ref{7}),  (\ref{7a})  can be efficiently calculated   within the INLA platform. In the next section, we briefly discuss how these calculations are carried within INLA.

\subsection{Bayesian inference using INLA}\label{inla}

Conditional on a realization of $W(s)$, a log-Gaussian Cox process is an inhomogeneous Poisson process. It follows that the likelihood for an LGCP is of the form

\begin{equation}
log({\cal F}(\theta|\bx))=|\Omega| - \int_\Omega \lambda_1(s|\bx,  \theta) ds + \sum_{s_i \in S} \lambda_1(s_i|\bx,\theta), \label{lik}
\end{equation}
where $S$ is the set of observed locations and $\lambda_1(s)$ is defined in (\ref{4}).

The integral in the likelihood is intractable due the stochastic nature of $\lambda_1(s)$. To solve this problem we could use the traditional methods to fit a log-Cox process, which consists of dividing the study regions into cells, forming a lattice, and then counting the number of points into each one. These counts are modeled using the Poisson likelihood. See for example Illian {\it et al} (2010). However, Simpson {\it et al} (2016) consider that this approach can be very inefficient, especially when the intensity of the process is high, the window of observation is too large or when the pattern is rare. They propose the use of an SPDE (Stochastic Partial Differential Equation) approach, introduced by Lindgren {\it et al} (2011), to transform a Gaussian field (GF) to a Gaussian Markov random field (GMRF). This methodology uses a computational mesh only for representing the latent Gaussian random field and not for modeling counts. Lindgren {\it et al} (2011) assume the following finite element representation

\begin{equation}
W(s) \approx \sum_{j=1}^{N} w_j \psi_j(s)
\end{equation}
where
$N$ is the number of the mesh nodes, $w=(w_1,w_2,...,w_N)^T$ is a multivariate Gaussian random vector (representing a Gaussian Markov random field, GMRF) and $\{\psi_j\}_{j=1}^N$ are the selected basis functions defined for each mesh node: $\psi_j$ is 1 at mesh node $j$ and 0 in all the other mesh nodes. $w$ is chosen in a way that the distribution of $W(s)$ approximates the distribution of the solution to an SPDE. Lindgren {\it et al} (2011) showed that the resulting distribution for the weights is $w \sim N(0, Q(\tau, k)^{-1})$ where the precision matrix $Q(\tau, k)$ is a polynomial in the parameters $\tau$ and $k$. Working directly with the SPDE parameters $k$ and $\tau$ can be difficult because they both affect the variance of the field (Yuan {\it et al} (2017)). So, we will consider the standard deviation $\sigma$ and the spatial range $\rho$ which are respectively given by

\begin{equation}
\sigma=\sqrt{\frac{1}{4\pi k^2 \tau^2}}
\end{equation}
and $\rho=\frac{\sqrt{8}}{k}$. After that approximation, it follows that the integral in (\ref{lik}) can be written as

\begin{equation}
\int_\Omega \lambda_1(s) ds=\int_\Omega \exp(W(s)) ds \approx \int_\Omega \exp \left(\sum_{j=1}^{N} w_j \psi_j(s) \right) ds
\end{equation}

This integral can be approximated using standard numerical integration schemes. Simpson {\it et al} (2016) suggest to use the follow quadrature rule

\begin{equation}
\int_\Omega f(s) ds \approx \sum_{i=1}^{N+n} \beta_i f(s_i)
\end{equation}
where $\{s_i\}_{i=1}^{N+n}$ are the locations of mesh nodes and observations, and $\{\beta_i\}_{i=1}^{N+n}$ are the quadrature weights.

Unlike the traditional methods  for  inference in LGCP models, this methodology uses each location to model the point pattern, without aggregation. The LGCP model  belongs to the latent Gaussian models and consequently, the inference can be done within the INLA platform.

As we noted, the SPDE methodology requires a triangulation of the study region to represent a Gaussian random field.  Here, we used a Delaunay triangulation with 3923 mesh nodes.

In real data applications, it is common that the point pattern and its associated marks are dependent. In our case, we expect that the average number of unemployed people per dwelling  to be dependent with the intensity of residential buildings, but the signal of that correlation is not obvious. On the one hand, we expect the number of unemployed people to be higher in regions with higher intensity of residential buildings and  on the other hand, we expected more opportunities of employment in these  regions.

Illian {\it et al} (2008) describes two  types of marked point process models depending on the type of dependence between the point patterns and marks. Here, we consider two versions of conditional dependence: In the first model we assume, as was explained in section \ref{spp},  that there is a common latent Gaussian field that govern the dependence structures of points and marks and conditional on this field, point patterns and marks are independent. In the second alternative model, we assume that there are two independent fields that govern the dependence structures of points patterns and marks. It is also possible to introduce a third coreginalization model (Banerjee {\it et al}, 2004, Gelfand {\it et al}, 2004) consisting of two dependent latent processes for point patterns and marks by assuming independence of points and marks conditional on these latent processes.   Coreginalization models can be inferred within the INLA platform, however this would require joint modeling of two dependent fields and we will not pursue these models in this work. In table \ref{DIC}, a comparison of these alternative models is given.

Here, we give the details of the model based on a common latent Gaussian model.
Let us consider that $\{s_i\}_{i=1}^{N+n}$ are the locations of the mesh nodes and the locations of the sampled residential buildings, and $\{y(s_i)\}_{i=1}^{N+n}$ are the number of unemployed people per residential building. The hierarchical structure of the model considered is given by

\begin{enumerate}

\item Data$|$Parameter

\begin{equation}
p(\{s_i,i=1,...,N+n\}|\lambda_1) \approx \prod_{i=1}^{N+n} \text{Poisson}(\beta_i \lambda_1(s_i))
\end{equation}
\begin{equation}
p(\{y(s_i),i=1,...,N+n\}| \lambda_2) \approx \prod_{i=1}^{N+n} \text{Poisson}(\lambda_2 (s_i))
\end{equation}
where $\beta_i$ is defined in (18).

\item Parameter$|$Hyperparameters

\begin{equation}
\log(\lambda_1(s_i))=\alpha_1+\text{offset}_1(s_i) +W(s_i), \label{101}
\end{equation}

\begin{equation}
\log(\lambda_2 (s_i))=\alpha_2+\text{offset}_2(s_i) +  Z_2'(s_i)\theta +\alpha_3 W(s_i), \label{102}
\end{equation}
where $W(s)$ is the GMRF given in (3).

\item Hyperparameters

\begin{equation}
\alpha_1 \sim N(0,1000)
\end{equation}
\begin{equation}
\alpha_2 \sim N(0,1000)
\end{equation}
\begin{equation}
\theta_j \sim N(0,1000), j=1,...,p
\end{equation}
\begin{equation}
\alpha_3 \sim N(0,1000)
\end{equation}

We assume that the latent field $W$ belong to the Matern class with $\nu=1$. We further assume that the model parameter of this field has the same prior structure as given below:

We followed Simpson {\it et al} (2017) and Fuglstad {\it et al} (2017) to construct a joint penalising complexity (PC) prior density for the spatial range, $\rho$, and the marginal standard deviation, $\sigma$, which is given by

\begin{equation}
p(\rho,\sigma)=R S \rho^{-2} e^{-R \rho^{-1} - S \sigma}
\end{equation}

where $R$ and $S$ are hyperparameters determined by $R=-\log(\alpha_1) \rho_0$ and $S=\frac{-\log(\alpha_2)}{\sigma_0}$.

The practical approach for this in INLA is to require the user to indirectly specify these hyperparameters through $P(\rho < \rho_0)=\alpha_1$ and $P(\sigma < \sigma_0)=\alpha_2$. Here, we considered $\rho_0=400, \alpha_1=0.5,  \sigma_0=1,\alpha_2=0.5$.

\end{enumerate}

The term  $\mbox{offset}_1 (s_i)$ in (\ref{101}) represents the log population density. We know their numbers by NUTS III regions so, based on that, we produced a spatial prediction for all domains by way of a Kernel smoothing, using the centroids of the NUTS III regions. The resulting prediction is given in figure \ref{offsets}. Lisboa, Porto and Península de Setúbal are the regions that stand out most. The $\mbox{offset}_2 (s_i)$  term in (\ref{102}) represents the log of the number of people per residential building.

We have the information for the residential buildings locations in the sample, but we need to estimate it for the mesh nodes. For this we also used the Kernel smoothing (see figure \ref{covariates2}).

\FloatBarrier

\subsubsection{Model selection}\label{mselection}

To evaluate the significance of each covariate and random effect in the marks, we considered  different models and compared the results of two model selection criteria: deviance information criterion ($DIC$) and Watanabe-Akaike information criterion ($WAIC$).

DIC, proposed by Spiegehalter {\it et al} (2002), is the most commonly used measure of model fit. It is based on a balance between the fit of the model to the data and the corresponding complexity of the model: $DIC = \bar{D} + p_D$ where $\bar{D}$ is the posterior mean deviance of the model and $p_D$ is the effective number of parameters. The model with the smallest value of $DIC$ is the one with a better balance between the model adjustment and complexity. However, this criterion can present some problems, which arise in part from not being fully Bayesian.

A typical alternative is the WAIC, proposed by Watanabe (2010), which is fully Bayesian in that it uses the entire posterior distribution. It can be considered as an improvement on the DIC for Bayesian models (Gelman \textit{et al}, 2014).

Several alternative spatial random effects were used in modelling the intensities, namely (i) Common random effect $W$ both for points and marks (ii) Random effect $W$ and its scaled version $\alpha_3 W$ for the points and marks respectively (iii) two independent latent processes $W$ and $W_2$  for the  points and their marks respectively.

Table \ref{DIC} shows the values of these two criterions for the models considered for the marked point process. In this case, the model with the best performance was the one that took into account the following factors:
\begin{itemize}
\item  the offset term given by the population density to model the intensity of the point process ($\text{offset}_{1}$) ;
     \item the covariates to model the mark intensity (number of individuals per residential building($nind_2$): the median of the education level ($\text{edu}_2$), the mean age ($\text{age}_2$), and the proportion of registered unemployed people ($\text{iefp}_2$). Here, subscripts $1$ and $2$ indicate that the corresponding covariate is used in modelling intensity $\lambda_1(s)$ and $\lambda_2(s)$  respectively;
         \item  two independent latent processes $W_1$ and $W_2$ used for points and their marks.
             \end{itemize}

\begin{table}[ht]
\caption{DIC,  WAIC and the effective number of parameters \label{DIC}}
%\centering
%\footnotesize{
%\tiny
\scalebox{0.8}{
\centering
\begin{tabular}{lrrrr}
\hline
log intensity of points and marks &  $DIC$ & $WAIC$ & $p_{DIC}$ & $p_{WAIC}$ \\
\hline
$\alpha_{1}  \: ;  \: \alpha_{2}$ &	110707.09 &  110724.00 & 2.18 & 19.04 \\
$\alpha_{1} + offset_{1} \: ;   \: \alpha_{2}$ &	84826.76 & 84853.04 & 2.198 & 28.41 \\
$\alpha_{1} + offset_{1}  \: ;  \: \alpha_{2} +  offset_2$ &	110707.09 & 110724.00  & 2.18 & 19.04\\
$\alpha_{1} + offset_{1}  \: ;   \: \alpha_{2}  + nind_2$ &	84791.91 & 84818.07  & 3.199 &29.29\\
$\alpha_{1} +  offset_{1}  \: ;  \: \alpha_{2}  + nind_2 + edu_2$ &	84790.11 & 84816.27  &4.198 &30.28\\
$\alpha_{1} + offset_{1}  \: ;  \: \alpha_{2} + nind_2 + edu_2 +age_2 $ &	84714.83 & 84741.04  &5.198 &31.34\\
$\alpha_{1} + offset_{1}  \: ;  \: \alpha_{2}  + nind_2 + edu_2 +age_2 + iefp_2$ &84706.69 & 84733.00  &6.197 &32.444\\
$\alpha_{1} + offset_{1} + W   \: ;  \:  \alpha_{2}  + nind_2 + edu_2 +age_2 + iefp_2 $ &	48538.24 & 67474.12  &1817.38 &15031.22\\
$\alpha_{1} + offset_{1} + W   \: ;   \: \alpha_{2} + nind_2 + edu_2 +age_2 + iefp_2  + \alpha_3 W$ &	48656.62 & 67562.53  & 1796.70 &14998.79\\
$\alpha_{1} + offset_{1}  + W_1   \: ;   \: \alpha_{2} + nind_2 + edu_2 +age_2 + iefp_2  + W_2$ &	48505.25 & 67443.79  &1842.59 &15058.96\\
\hline
\end{tabular}
%\caption{DIC,  WAIC and the effective number of parameters \label{DIC}}
}
\end{table}

Here, $p_{\text{DIC}}$ and $p_{\text{WAIC}}$ are the effective number of parameters, as described in Spiegelhalter {\it et al} (2002) and Gelman {\it et al} (2014), respectively.

It is clear from the table that the model that employs all the covariate information and two independent latent processes, one for points other for marks seems to give the best fit with the model that employs  all the covariate information and a single common latent process for the points and marks coming second. Here, we chose the model with lower DIC to continue with these analysis. We also considered a  negative binomial distribution for the marks as an alternative to the poisson marks, but these models did not bring gain in terms of DIC.

\begin{figure}[!htbp]
\centering
\includegraphics[width=9cm]{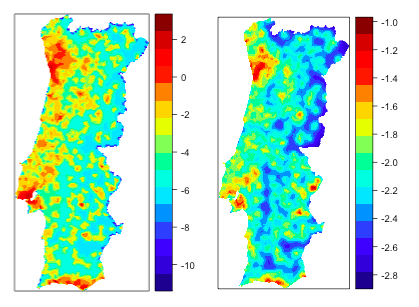}
\caption{Posterior mean of: log intensity (left), and log mean marks (right)}
\label{map_points}
\end{figure}

To perform the spatial prediction,  we created a regular grid of 1$\text{km}^2$ in the domain. A projection from the mesh to the grid was performed and the resultant maps of the posterior mean of the logarithmic transformation of the intensity of the residential units $\log(\lambda_1(s))$  and the logarithmic of the marks mean are shown in figure \ref{map_points}. The plot of the logarithmic transformation of the intensity provides a clearer image about the spatial variation of the residential buildings. As we expected, the highest values are concentrated in Lisboa, Porto, and Algarve regions. The intensity is clearly higher near the coast and lower in the interior of the country.

\begin{figure}[!htbp]
\centering
\includegraphics[width=9cm]{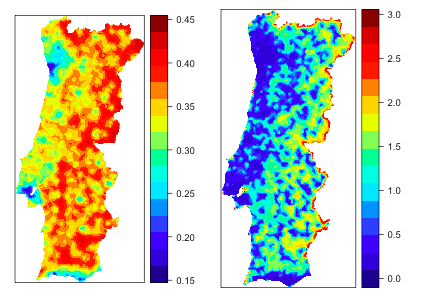}
\caption{Standard deviation of: log intensity (left), and log mean marks (right)}
\label{map_sd}
\end{figure}

The standard deviations of these fields are plotted in figure \ref{map_sd}.

With these estimates, we can conclude that the average number of unemployed people per residential building is higher in the Grande Porto, Península de Setúbal and Alentejo Central regions.

\FloatBarrier

\subsection{Model validation}

\begin{figure}[!htbp]
\centering
\includegraphics[width=8cm]{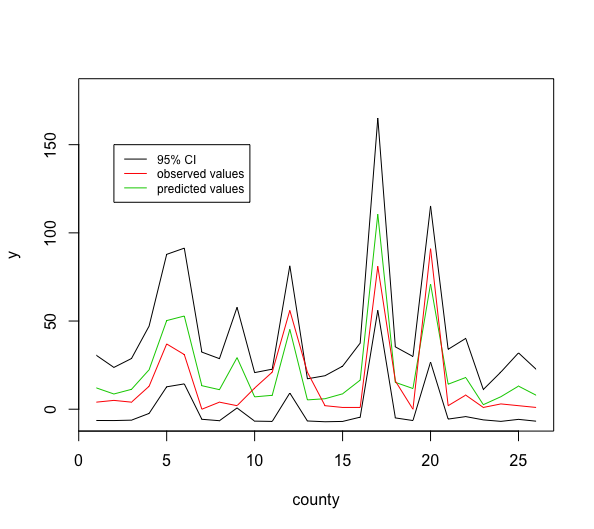}
\caption{95\% CI, observed values and predicted values for the 26 counties that were removed from the sample}
\label{mv1}
\end{figure}

\begin{figure}[!htbp]
\centering
\includegraphics[width=8cm]{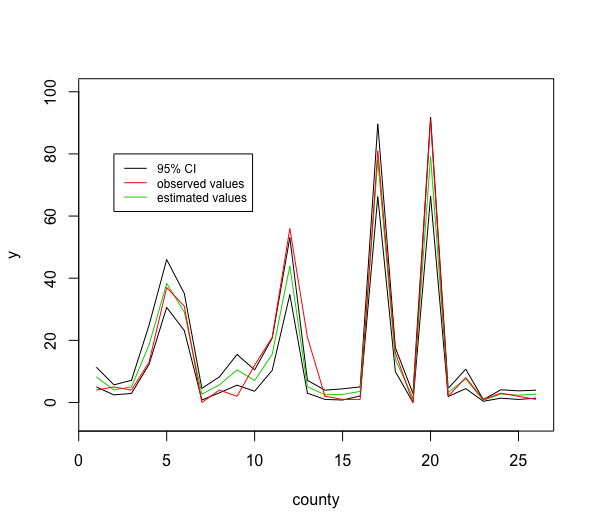}
\caption{95\% CI, observed values and predicted values for the 26 counties that were removed from the sample for the first plot. Here, we used all counties in the sample for the modeling process.}
\label{mv6}
\end{figure}

\begin{figure}[!htbp]
\centering
\includegraphics[width=8cm]{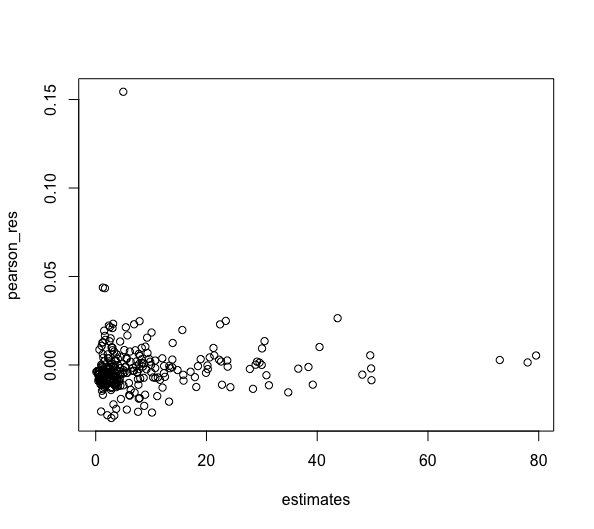}
\caption{Pearson residuals versus fitted values for the 278 counties}
\label{mv9}
\end{figure}

\begin{figure}[!htbp]
\centering
\includegraphics[width=8cm]{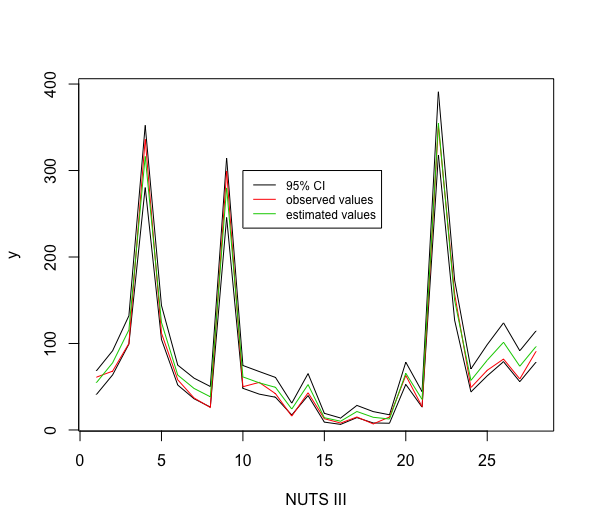}
\caption{95\% CI, observed values and fitted values for the 28 NUTS III regions}
\label{mv10}
\end{figure}

For model validation purposes, we chose randomly 26 counties and fitted the model excluding the data from these counties. Figure \ref{mv1} gives the 95\% credible intervals for the predicted values of unemployment from the model together with the observations and predicted values for these 26 counties.  Figure \ref{mv6} gives the credible intervals together with observations and estimates for the same 26 chosen counties when all the data are used in fitting the model.   Figure \ref{mv9} gives the  Pearson residuals versus fitted values for the 278 counties. Figures \ref{mv10} gives the 95\% credible intervals, observations and estimates for the NUTS III regions. As is expected, the model gives higher precision estimates at NUTS III regions as compared to county level.

\FloatBarrier

\subsection{Unemployment estimation for NUTS III regions and counties}\label{thinning}

The marked point process model explained in the previous section, projects the sampling survey in space. This  point process is a thinned version of the true point patterns of the residential units together with their  marks across Portugal.

Let $N_1^*(s)$ represent  the true point patterns of the residential units with intensity $\lambda_1^*(s)$.
Then
\begin{equation}
\lambda_1^*(s)=\frac{\lambda_1(s)}{P(RU(s))}, \label{t1}
\end{equation}
where, $P(RU(s))$ is the probability that a residential unit at $s$ is included in the survey. $P(RU(s))$ should be interpreted as  the proportion of the residential units in any infinitesimal area which is included in the sampling survey.
Assume also that $N_2^*(s)$ represent the true intensity of the  number of unemployed observed in residential unit at location $s$. then the intensity $\lambda_2^*(s)$  of this counting process is given by
\begin{equation}
\lambda_2^*(s)=\frac{\lambda_2(s)}{P(D(s)|RU(s))}, \label{t2}
\end{equation}
where the probability  $P(D(s)|RU(s))$  should be interpreted as the proportion of dwellings in a residential unit which are included in the sampling survey.

Target quantities $\ref{7}$ and $\ref{7a}$   depend on the multiplicative intensity
$\lambda_1(s)\lambda_2(s)$, which is a thinned version of $\lambda_1^*(s)\lambda_2^*(s)$ and this relationship is given by

\begin{equation}
\lambda_1^*(s)\lambda_2^*(s)=\frac{\lambda_1(s)\lambda_2(s)}{p(s)}, \label{t3}
\end{equation}

where
\begin{eqnarray}
p(s)&=& P(RU(s))P(D(s)|RU(s))\nonumber\\
&=& P(D(s)),\nonumber\end{eqnarray}

since $P(D(s)|RU^c(s))=0$.
Here, $p(s)$ should be interpreted as the proportion of dwellings that are chosen in the sampling survey. As explained in section \ref{survey}  these probabilities are  estimated using (\ref{thinning2}).

To define the intensity of the full version of the spatial point process, the knowledge of the sampling probabilities $p(s)$ for whole domain  is required. Here, we estimate these probabilities using the kernel method, based on the data given by the sampling survey. This method allows us to generate a spatial prediction for the centers of the cells of the grid, derived from the values of the dwellingss locations.
%and these estimates reveal a considerable difference between the north and south of the country as shown in figure \ref{prob}.

%\begin{figure}[!htbp]
%\centering
%\includegraphics[width=0.35\linewidth]{imagens_novas/kernel_prob_sel_a.png}
%\caption{Kernel estimates of the selection probabilities of the dwellings.}
%\label{prob}
%\end{figure}

%Target quantities (\ref{7}) and (\ref{7a}) can easily be adjusted to the full version by substituting $\lambda_1(s)$ by $\lambda^*_1(s)$ given in (\ref{thin}).

We simulated 1000 values of the predictive posterior distribution of $\lambda_1$ and $\lambda_2$ for each cell $I_{j}$ to estimate the target quantities, by simulating samples from the posterior distributions of the model parameters and the latent gaussian markov fields used in the model.

\begin{figure}[!htbp]
\centering
\includegraphics[width=6.5cm]{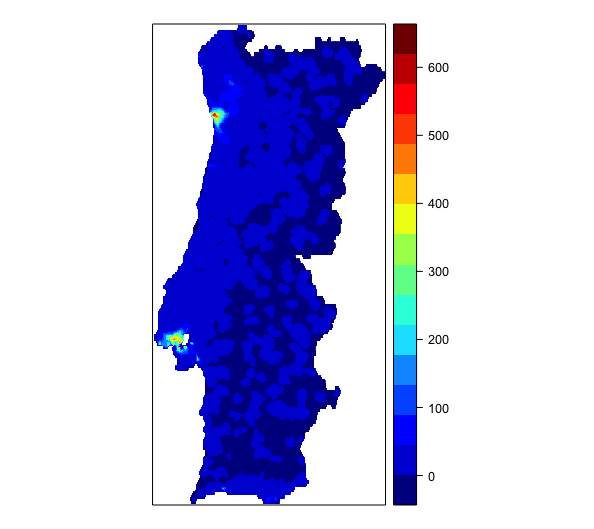}
\caption{Predictive multiplicative intensity function}
\label{multiplicative}
\end{figure}

Figure \ref{multiplicative} gives the predictive multiplicative intensity function
\begin{equation}
\frac{\lambda_1(s|\bx)\lambda_2(s|\bx)}{p(s)}, \label{pi}
\end{equation}
which will form the basis for calculating the unemployment in any region $A$, expressed in terms of $E(N(A)|\bx)$ as given in \ref{7}.

\begin{figure}[!htbp]
\centering
\includegraphics[width=12cm]{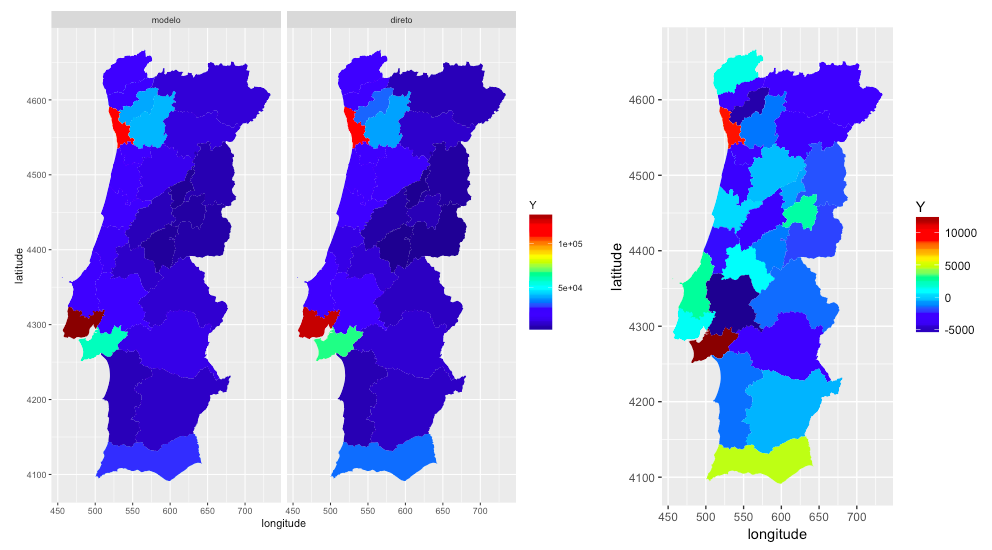}
\caption{Comparison between model and direct estimates by NUTS III regions}
\label{model_direct}
\end{figure}

 Figure \ref{model_direct} shows the estimates at NUTS III level obtained by this model and the estimates obtained by the direct method, as well as the differences. Indeed, the spatial distribution is very similar in both methods, as is expected.

%There is a clear discrepance in the variation coefficients obtained by these methods (see figure \ref{map_cvs}). This map remit to the problem of the small area estimation, where the estimates are not reliable for small areas (areas with low population density). Indeed, in these areas we can see that point referenced method we propose  has a better performance in terms of variation coefficients.

%Figure \ref{var_model_direct} gives the comparison of variance of estimates for NUTS III region obtained from the direct estimation method and  $Var_{(W(s), \btheta|\bx)}\left[E(N(A)|\bx,W(s), \btheta)\right]$, with $A$ representing each of the NUTS III regions. Although big differences are not visible due the differences of the estimates values between the regions, there is a clear discrepance in the variation coefficients obtained by these methods (see figure \ref{map_cvs}). This map remit to the problem of the small area estimation, where the estimates are not reliable for small areas (areas with low population density). Indeed, in these areas we can see that point referenced method we propose  has a better performance in terms of variation coefficients.

\begin{figure}[!htbp]
\centering
\includegraphics[width=8cm]{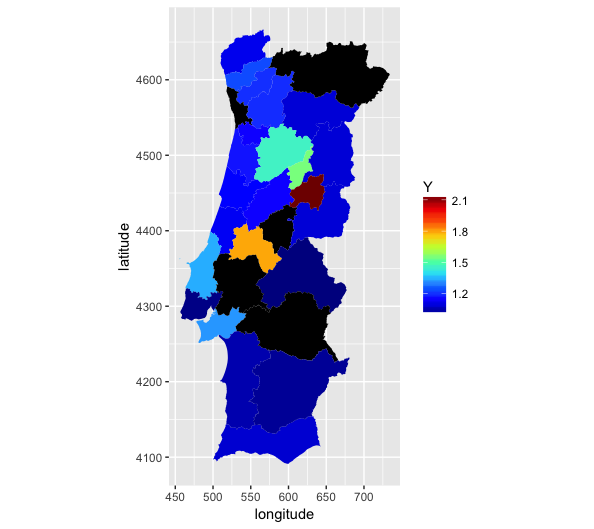}
\caption{Ratio between the standard deviation of the direct method and model estimates. Black scales indicate NUTS III regions where the ratio is less than 1.}
\label{var_model_direct}
\end{figure}

Figure \ref{var_model_direct} gives the ratio between the standard deviation of estimates obtained from the direct estimation method and square root of  $\text{Var}_{(W(s), \theta|\bx)}\left[E(N(A)|\bx,W(s), \theta)\right]$,  with $A$ representing each of the NUTS III regions. As we can see, the ratio is higher than 1 for the most part of regions, indicating clearly that point referenced method produces estimates with higher precision.

\begin{figure}[!htbp]
\centering
\includegraphics[width=8cm]{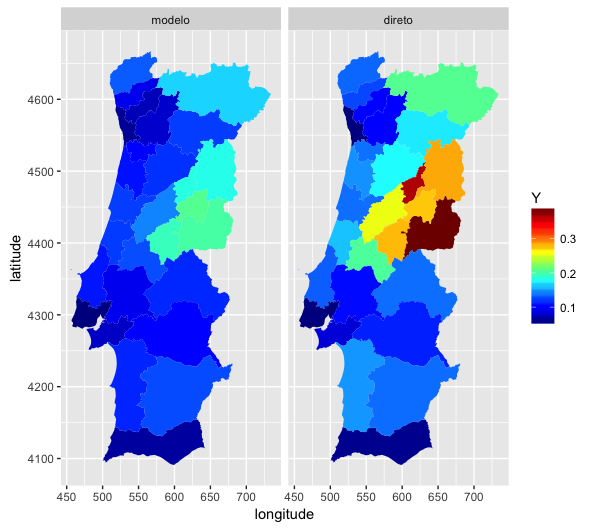}
\caption{Comparison of the coefficients of variation between the model applied and the direct method}
\label{map_cvs}
\end{figure}

Coefficients of variation, given by the ratio of the standard deviation to the mean, were also calculated and there is a clear discrepancy in the methods (see figure \ref{map_cvs}). This map remits to the problem of the small area estimation, where the estimates are not reliable for small areas (areas with low population density). Indeed, in these areas we can see that point referenced method we propose  has  better performance in terms of coefficients of variation.

\begin{figure}[!htbp]
\centering
\includegraphics[width=10cm]{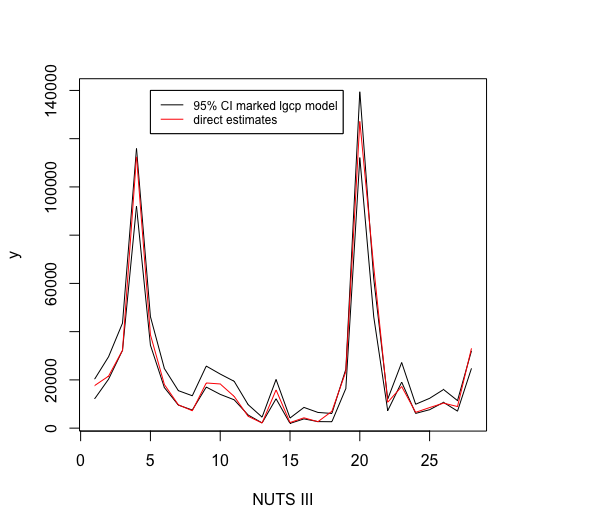}
\caption{95\% credible intervals for the posterior mean (black) and direct estimates (red) by NUTS III regions}
\label{ci}
\end{figure}

We also calculated the credible intervals for the posterior mean by NUTS III and compared those with the direct estimates (figure \ref{ci}). We note that the direct estimates are inside the intervals for almost all regions. The highest points correspond to Porto and Lisboa. The figure reveals that there is an underestimation in these regions and an overestimation in the others. This behaviour is probably due to the large differences in the intensity in  these regions and consequent smoothing of intensities across the space.

\begin{figure}[!htbp]
\centering
\includegraphics[width=14cm]{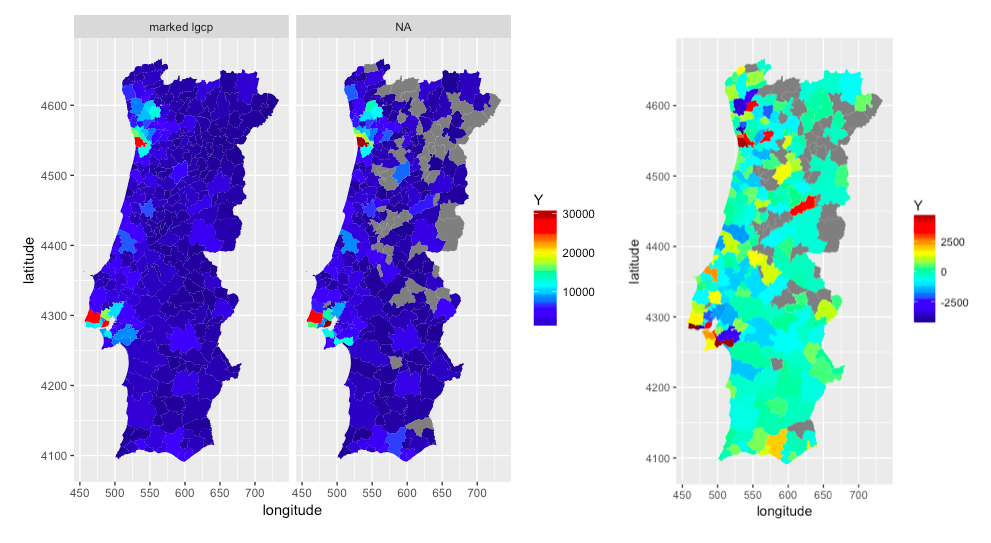}
\caption{Model and direct estimates, and their differences by county. Gray scales indicate counties where it is not possible to get direct estimates.}
\label{model_direct_concelho}
\end{figure}

We also give estimation results on a higher resolution, namely for 278 counties of mainland Portugal. Figure \ref{model_direct_concelho} gives comparative results from our model and the direct estimation method. Further comparisons on the standard deviations and coefficients of variation clearly  indicate that   point referenced methods we employ give estimates with higher precision at higher resolutions, as expected.

\FloatBarrier

\section{Discussion,  conclusions and further work }\label{discussion}

In this study, we employed point referenced spatial models to unemployement estimation  making use of  the  newly available  geo-referenced  quarterly sample surveys. Unlike the direct method and area level models, our approach  take into account specific information about the families including their spatial location at a very detailed geographical level. Therefore, it is natural that both the methodologies produce estimates at NUTS III level with higher precision than the direct method.

It is interesting to see that after applying these complex methodologies the results do not seem to be very different than those obtained by the direct method at  NUTS III level. Despite this similarity, these  methodologies provide important advantages in the small area and official statistics context, particularly delivering  reliable estimates also for much smaller areas. As opposed to  areal models proposed in the literature, these models take into account of  the sampling design, which is important in this context.

National Statistical Institutes usually require a great deal of consistency between the estimates obtained at different areal resolutions, but this requirement is not easy to satisfy using the standard small area models. Since this new methodology operates independently from the administrative limits of geographical units, it can provide us with the necessary means to meet this requirement both a consistent and timely fashion.

Despite the complexity of this methodology, the computational costs are not high due to the availability of the R-INLA package in the software R.

In this work, we do not report on the time dynamics of unemployement. At present, there are 14 quarterly sample surveys with geo-referenced sampling units. As these quarterly data further become available,  it may be possible to investigate how spatial variation of unemployment changes over time. This can be done by  considering space-time marked point processes, in which the latent  process now is a space time Gaussian process.  and by adding  time varying covariates in the model such as a linear or quadratic trends functions in time. It is possible to infer on such models within the INLA platform.
Recently, INE has started on a much more ambitious  geo-referencing  methods by identifying and geo-referencing all the residential units in Portugal. Methods and models that are adequate for these new realities  will be discussed elsewhere as new data become available.

We expect that this methodology will be even more beneficial when time-dynamics will be incorporated, as it is more intuitive that the current results today would act as the prior for the next time point and, additionally, there is a benefit of doing some smoothing in time.

\section*{Acknowledgements}

This work was supported by the project $UID/MAT/00006/2013$ and the PhD scholarship $SFRH/BD/92728/2013$ from Fundação para a Ciência e Tecnologia. Instituto Nacional de Estatística and Centro de Estatística e Aplicações da Universidade de Lisboa are the reception institutions. We would like to thank professor Antónia Turkman, Elias Krainski and Paula Pereira for their help.

\section*{Note}

This study is the responsibility of the authors and does not reflect the official opinions of Instituto Nacional de Estatística.

\FloatBarrier

\end{document}